\documentclass[twocolumn,showpacs,preprintnumbers,amsmath,amssymb,
floatfix,nofootinbib]{revtex4}
\usepackage{graphicx}
\usepackage{dcolumn}
\usepackage{bm}
\usepackage{amsthm}
\usepackage{amsmath}
\usepackage{amsfonts}
\usepackage{color}

\newcommand{\be}{\begin{equation}}
\newcommand{\ee}{\end{equation}}
\newcommand{\ba}{\begin{eqnarray}}
\newcommand{\ea}{\end{eqnarray}}

\newcommand{\n}[1]{\label{#1}}
\newcommand{\eq}[1]{(\ref{#1})}
\newcommand{\pa}{\partial}

\newcommand{\hh}{\, ,\hspace{0.5cm}}
\newcommand{\hhh}{\, ,\hspace{0.2cm}}

\begin{document}

{\hspace{15cm} \mbox{YITP--11--90}}

\title{Weakly magnetized black holes as particle accelerators}
\author{Valeri P. Frolov}
\email{vfrolov@ualberta.ca}
\affiliation{Theoretical Physics Institute, University of Alberta,
Edmonton, AB, Canada,  T6G 2G7, and
Yukawa Institute for Theoretical Physics, Kyoto University, Kyoto, Japan}
\date{\today}

\begin{abstract}
We study collision of particles in the vicinity of a horizon of a weakly magnetized non-rotating black hole. In the presence of the magnetic field innermost stable circular orbits (ISCO) of charged particles can be located close to the horizon. We demonstrate that for a collision of two particles, one of which is charged and revolving at ISCO and the other is neutral and falling from infinity, the maximal collision energy  can be  high
in the limit of strong magnetic field. This effect has some similarity with the recently discussed effect of high center-of-mass energy for collision of particles in extremely rotating black holes. We also demonstrate that for `realistic' astrophysical black holes their ability to play the role of `accelerators' is in fact quite restricted.
\end{abstract}

\pacs{04.70.Bw, 04.70.-s, 04.25.-g }

\maketitle

\section{INTRODUCTION}

Collision of particles near a rotating black hole horizon may produce high energy radiation. Such processes connected with the Penrose effect have been studied long time ago by Piran and collaborators
\cite{Piran:1975,Piran:1977dm,Piran:1977aa}.
Recently Ba\~{n}ados, Silk and West (BSW) \cite{Banados:2009pr}
demonstrated that for an extremely rotating black hole such collisions can produce particles of  high center-of-mass energy. In an idealized set up this energy can be higher than Planckian energy, so that one might think about  black holes as super high energy colliders. This work stimulated a lot of interest to this problem
\cite{Banados:2010kn,Williams:2011uz,Harada:2010yv,Kimura:2010qy,Harada:2011xz,Harada:2011pg,Zaslavskii:2010pw,
Zaslavskii:2011dz,Zaslavskii:2011uu}. More detailed analysis demonstrated that this model is oversimplified. There exist several processes which suppress the possible high value of the collision energy \cite{Jacobson:2009zg,Berti:2009bk}. The center-of-mass energy can be infinitely high only if the black hole is extremely rotating. A tiny violation of the extremality condition, which is practically inevitable in the astrophysical applications,  strongly suppresses this effect. Even in the idealized extremely rotating case special fine-tuning is required for particle trajectories. The gravitational radiation also significantly affects this process.

The aim of this paper is to show that a similar effect of particle collision with high center-of-mass energy is also possible when a black hole is non-rotating (or slowly rotating) provided there exists magnetic field in its exterior.
This observation might be interesting since
there exist both theoretical and experimental indications that such a
magnetic field must be present in the vicinity of black holes.
In what follows we assume that this field is weak and its energy-momentum does not modify the background black hole geometry.
For a black hole of mass $M$ this condition holds if the strength of magnetic field satisfies the condition
\be\n{BBB}
B\ll B_{max}={c^4\over G^{3/2}
M_{\odot}}\left(\frac{M_{\odot}}{M}\right)\sim
10^{19}{M_{\odot}\over M}\mbox{Gauss}\, .
\ee
We call such black holes {\em weakly magnetized}. One can expect that the condition \eq{BBB} is satisfied both for stellar mass and supermassive black holes.

The condition \eq{BBB} does not mean that the magnetic field does not affect the charged particle motion. The following dimensionless quantity
\be
b={|q|B GM\over m c^4}\,
\ee
can be used to characterize relative strength of magnetic and gravitational forces acting on a charged particle (with charge q and mass $m$) in the vicinity of the weakly magnetized black hole of mass $M$. For electrons this quantity is
\be\n{bdef}
b \sim 8.6 \times 10^{10} \left({B\over 10^8 \mbox{Gauss}}\right) \left({M\over 10 M_{\odot}}\right) \, ,
\ee
while for protons it is less by the factor $m_p/m_e\approx 1836$.
If one uses the following estimations for the magnetic field in the vicinity of black hole, presented in \cite{MF},
\ba\n{BBest}
B&\sim& 10^8 \mbox{Gauss     } \mbox{for   }  M\sim 10 M_{\odot} \mbox{   black holes;}\n{BBbesta}\\
B&\sim& 10^4 \mbox{Gauss     } \mbox{for   }  M\sim 10^{9} M_{\odot} \mbox{   black holes}\, ,\n{BBbestb}
\ea
one obtains $b\sim 8.6 \times 10^{10}$ and $b\sim 8.6 \times 10^{14}$ for these two cases, respectively. This means that even for weakly magnetized black holes the magnetic field can dramatically modify the charged particle motion. For $B=b_{max}$ the parameter $b$ reaches the value $b_{max}=8.6\times 10^{20}$, which is independent of the mass of a black hole.

It is interesting that in some aspects the action of the magnetic field on charged particle motion is similar to the effects of the rotation of the black hole on the motion of neutral particles in its vicinity. Such a similarity can be expected in a general case because of the well known gravito- electromagnetism analogy (see e.g. \cite{MGL,BM}).
In particular, in the presence of the magnetic field the innermost stable circular orbit (ISCO) for a charged particle can be arbitrary close to the horizon \cite{GP}, as it occurs for an extremely rotating black hole.  For this reason one can expect that under special conditions magnetized black holes can play a role of accelerators, similar to the rotating ones. The purpose of this paper is to demonstrate that this is really so.

Many important results concerning charged particle motion in magnetized can be found in \cite{GP,AG,AO}. A more recent paper \cite{FS} contains more detailed study of this problem. The results of the latter paper will be used in the present work.

The present paper is organized as follows. Section~II contains discussion of charged particle motion in the magnetized black holes, with the main focus on the properties of their ISCOs. In Section~III we discuss particles collisions in the vicinity of a weakly magnetized black hole. We demonstrate that for collision of two charged particles, moving along ISCO in the opposite directions, the center-mass energy is only slightly more that $2m$. However the collision energy can be high in the other case, when a freely falling neutral particle collides with  a charged particle at ISCO. Formally, the center-mass energy can be arbitrary high for this case, as it happens in the extremely rotating black holes. In Section~IV we discuss the obtained results. In particular, we show that the dependence of maximal collision energy ${\cal M}$ on the magnetic field parameter $b$ has the form ${\cal M}\sim b^{1/4}$. As a result one cannot reach extremely high energy ${\cal M}$ for realistic magnetic fields. We also discuss other effects which might be important for such collision processes.

\section{Charged particle motion in weakly magnetized black holes}

Let us discuss first a charged particle motion in the vicinity of a
Schwarzschild black hole of mass $M$ in the presence of an
external static axisymmetric and uniform at the spatial infinity magnetic field.
The Schwarzschild metric reads
\be\n{5}
ds^2=-fdt^2+f^{-1}dr^2+r^2 d\omega^2
\hh
f=1-\frac{r_g}{r}\, ,
\ee
where $r_g=2GM$ and $d\omega^2=d\theta^2+\sin^2\theta d\phi^2$. The
commuting Killing vectors ${\bm \xi}_{(t)}=\pa/\pa t$ and
${\bm \xi}_{(\phi)}=\pa/\pa \phi$   generate time translations and
rotations around the symmetry axis, respectively. The magnetic field potential in the Lorentz gauge
$A^{\mu}_{\,\,\,;\mu}=0$ is of the form
\be\n{A}
A^{\mu}={B\over 2}\xi^{\mu}_{(\phi)}\, .
\ee
The corresponding magnetic field is homogeneous at the spatial
infinity where it has the strength $B$ (see, e.g., \cite{AG,Wald}).
In what follows, we assume that $B\geq0$.

Dynamical equation for a charged particle motion is
\be
m{du^{\mu}\over d\tau}=qF^{\mu}_{\,\,\,\nu}\,u^{\nu}\,,\n{1}
\ee
where $\tau$ is the proper time, $u^{\mu}$ is the particle 4-velocity, $u^{\mu}u_{\mu}=-1$,
$q$ and $m$ are its charge and mass, respectively.
For the motion in the magnetized black hole  there
exist two conserved quantities associated with the Killing vectors:
the energy $E>0$ and the generalized azimuthal angular momentum
$L\in(-\infty,+\infty)$,
\ba
&&E\equiv -\xi^{\mu}_{(t)}P_{\mu}=m\,{dt\over d\tau}\left(1-\frac{r_g}{r}\right)\,,\n{8}\\
&&L\equiv \xi^{\mu}_{(\phi)}P_{\mu}=\left(m\,{d\phi\over d\tau}
+\frac{1}{2}qB\right) r^2\sin^2\theta\,.\n{9}
\ea
Here $P_{\mu}=m\,u_{\mu}+qA_{\mu}$ is the generalized 4-momentum of
the particle. We focus on the motion of the charged particle in the
equatorial plane, so that these integrals of motion are sufficient for
the completely integrability of the equations of motion\footnote{
Let us notice that the black hole parameters do not enter the expression
for the generalized azimuthal angular momentum, so that expression (\ref{9})
is the same as in a flat spacetime. If a particle does not move and is located at
$\theta=0$, \eq{9} takes the form $L=\frac{1}{2}qB r^2$. This means that
$L$ depends on the choice of the center of the spherical coordinates. This ambiguity
reflects the fact that $L$ is not a gauge invariant quantity, and by a proper gauge
transformation its value at a initial moment of time can be changed. For the
same reason the absolute value of the constant $L$ is not so important for
our problem. However, in the case of a black hole it is convenient to fix
the origin of the spherical coordinates in the `center' of the black hole, and to choose
the gauge in which \eq{A} is valid.
}.

It is convenient to use the following  dimensionless versions of $r$,  $\tau$, $E$,  $L$ and $B$
\be
\rho={r\over r_g},\, \,  \sigma={\tau\over r_g},\, \,
\ell={L\over mr_g},\, \,  {\cal E}={E\over m},\, \,  b={qBr_g\over 2m}\, .
\ee
Written in the dimensionless form the equations of motion are
\ba
&&\left(\frac{d\rho}{d\sigma}\right)^2={\cal E}^2-U\hh {d{\cal T}\over d\sigma}=\frac{{\cal E}\rho}{\rho-1}\,,\n{16}\\
&&\rho \frac{d\phi}{d\sigma}=\beta\hh \beta=\frac{\ell}{\rho}-b\rho\,,\n{17}
\ea
where the effective potential is
\be\n{22}
U=\left(1-\frac{1}{\rho}\right) (1+\beta^2)\,.
\ee
In what follows we assume that the charge $q$ is positive, so that $b$ is positive as well. 
We discuss only such solutions. Solutions for the negative charge can be obtained  by a 
simple substitution $b\to -b$, $\ell\to -\ell$, and $\phi\to -\phi$

We consider circular motion of the charged particle. The effective potential has minimum at the radius of such an orbit.  The momentum of a particle at the circular orbit of radius $r$ is
\ba
p^{\mu}&=&m \gamma (e_{(t)}^{\mu}+v e_{(\phi)}^{\mu})\, ,\n{pp}\\
e_{(t)}^{\mu}&=&f^{-1/2}\xi_{(t)}^{\mu}=f^{-1/2}\delta_t^{\mu}\, ,\\
e_{(\phi)}^{\mu}&=&r^{-1}\xi_{(\phi)}^{\mu}=r^{-1}\delta_{\phi}^{\mu}\, .
\ea
Here $v$ (which can be both positive and negative) is a velocity of the particle with respect to a rest frame, and
$\gamma$ is the Lorentz gamma factor. From normalization condition $\bm{p}^2=-m^2$
one has $\gamma=(1-v^2)^{-1/2}$. For $q>0$ the Lorentz force acting on a particle with $v>$ is {\it repulsive} (i.e. directed outwards the black hole), while for $v<0$ it is {\it attractive}.

Using relation $d\phi/d\tau=v\gamma/r$ and \eq{9} one gets
\be
v\gamma=\beta\, .
\ee
This relation allows one to find
\be\n{gvb}
\gamma^2=1+\beta^2\hh
v={\beta\over \sqrt{1+\beta^2}}\, .
\ee

The position of the innermost stable circular orbit (ISCO) is determined by the equations $\partial_{\rho} U=\partial^2_{\rho} U=0$. These conditions give 2 relations for 3 quantities, $\rho$, $\ell$, and $b$.
In the absence of the magnetic field the ISCO radius  is the same for both types of the directions of the motion and the corresponding value of $\rho$ is $\rho_{\pm}=3$.  For non-vanishing magnetic field the radii of the ISCO for positive and negative $\ell$ are  different. Both $\rho_{\pm}$ are smaller than $3$, and one  always has $\rho_+<\rho_-$.

\begin{figure}[htb]
\begin{center}
\includegraphics[width=5cm]{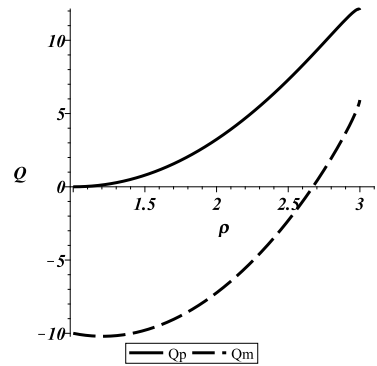}
\caption{$Q_+$ and $Q_-$ as functions of $\rho$.}\label{fig_1}
\end{center}
\end{figure}

Using equations (57) and (58) of the paper \cite{FS} one obtains
\ba\n{beta}
\beta_{\pm}&\equiv &{\ell_{\pm}\over \rho_{\pm}}-b_{\pm}\rho_{\pm}
=\pm {1\over \sqrt{2}}{\sqrt{3\rho_{\pm}-1}\mp \sqrt{3-\rho_{\pm}}\over Q_{\pm}^{1/2}}\nonumber\\
Q_{\pm}&=&4\rho_{\pm}^2-9\rho_{\pm}+3\pm A_{\pm}\, ,\nonumber\\
A_{\pm}&=&\sqrt{(3\rho_{\pm}-1)(3-\rho_{\pm})}\, .
\ea
Here $\rho_{\pm}$ is the radius of ISCO . For both types of motion $A_{\pm}$ is real only in the interval $1/3\le \rho\le 3$. The function $Q_+=0$ for $\rho_+=1$ and $Q_-=0$ for $\rho_-=\rho_{-,min}\equiv (5+\sqrt{13})/4$. Figure~\ref{fig_1} shows that the regions allowed for the ISCO radii are $1<\rho_+\le 3$ and $\rho_{-,min} < \rho_-\le 3$.

\begin{figure}[htb]
\begin{center}
\includegraphics[width=5cm]{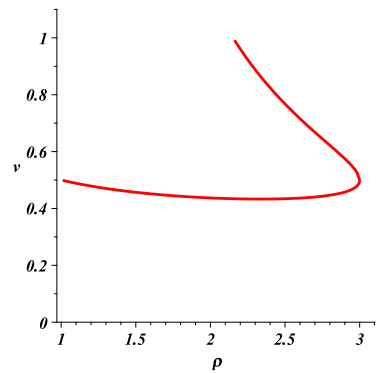}
\caption{Velocity of a particle at the  ISCO as a function of  its radius.}\label{fig_0}
\end{center}
\end{figure}

Using \eq{gvb} one finds the values of the velocity $v_{\pm}$ and $\gamma$-factor $\gamma_{\pm}$. Figure~\ref{fig_0} shows the velocity of a particle at the  ISCO as a function of  its radius.
The lower brunch, which starts at $\rho_+=1$ at the value $1/2$, is slightly less than $1/2$ in the interval  $(1,3)$. At the end of this interval it reaches the value $1/2$ again. The upper brunch starts with the same value $1/2$ at $\rho_-=3$ and monotonically increases until it reaches the value $1$ at $\rho=\rho_{-,min}$. Thus the motion of the particle at the ISCO becomes ultra-relativistic only in this limit.

One can write expression for $\gamma_{\pm}$ in the following form
\be
\gamma_{\pm}={2(\rho_{\pm}-1)\over [4\rho_{\pm}^2-9\rho_{\pm}+3\pm A_{\pm}]^{1/2}}\, .
\ee
The value of the effective potential $U$ at the position of the ISCO is
\be
U_{\pm}=(1-\rho_{\pm}^{-1})(1+\beta_{\pm}^2)\, .
\ee

\begin{figure}[htb]
\begin{center}
\includegraphics[width=6cm]{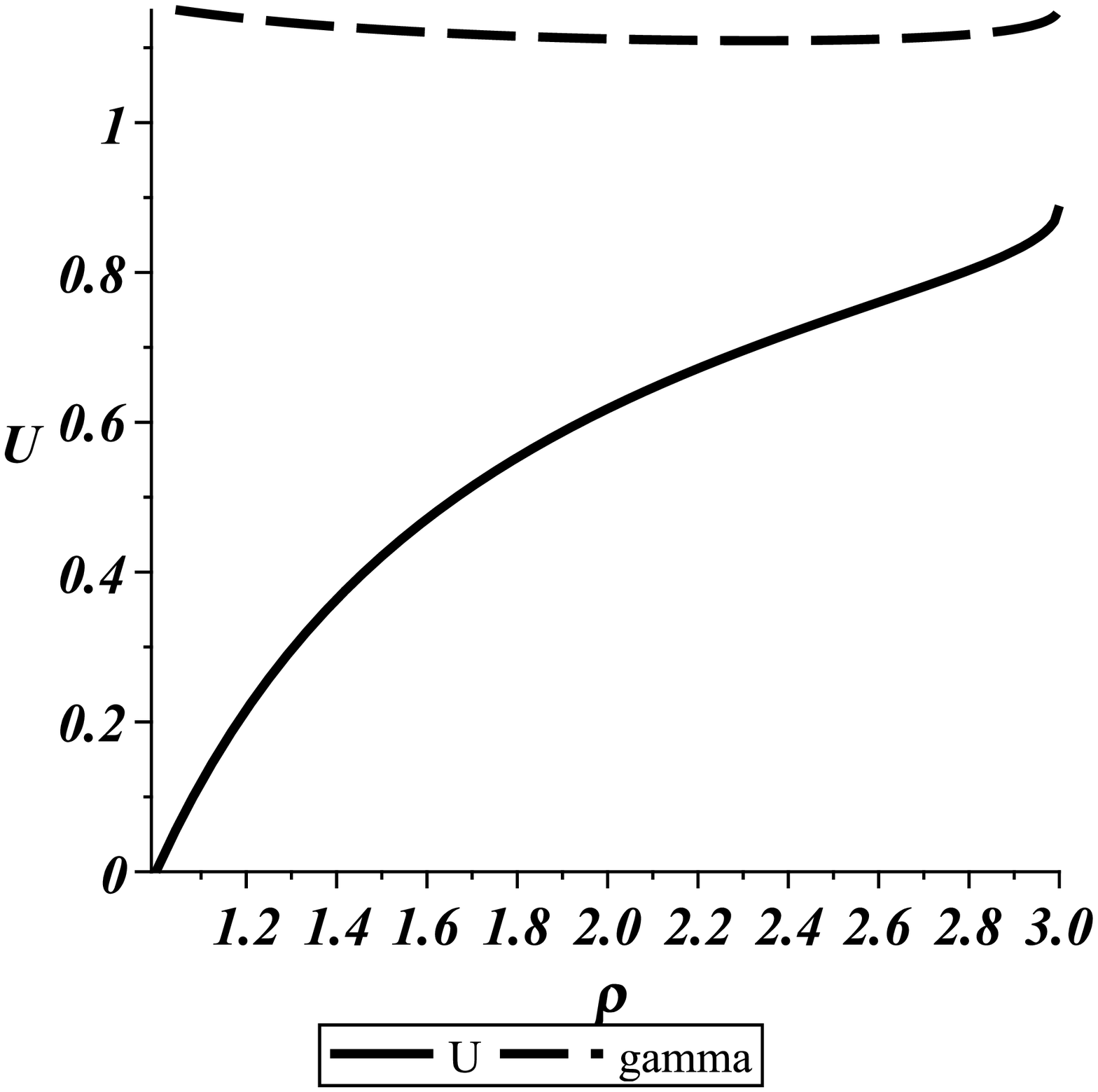}
\caption{The value of the effective potential $U_+$ at the position of the ISCO $\rho_+$ and the corresponding value of the $\gamma$-factor $\gamma_+$.}\label{fig_2p}
\end{center}
\end{figure}

\begin{figure}[htb]
\begin{center}
\includegraphics[width=6cm]{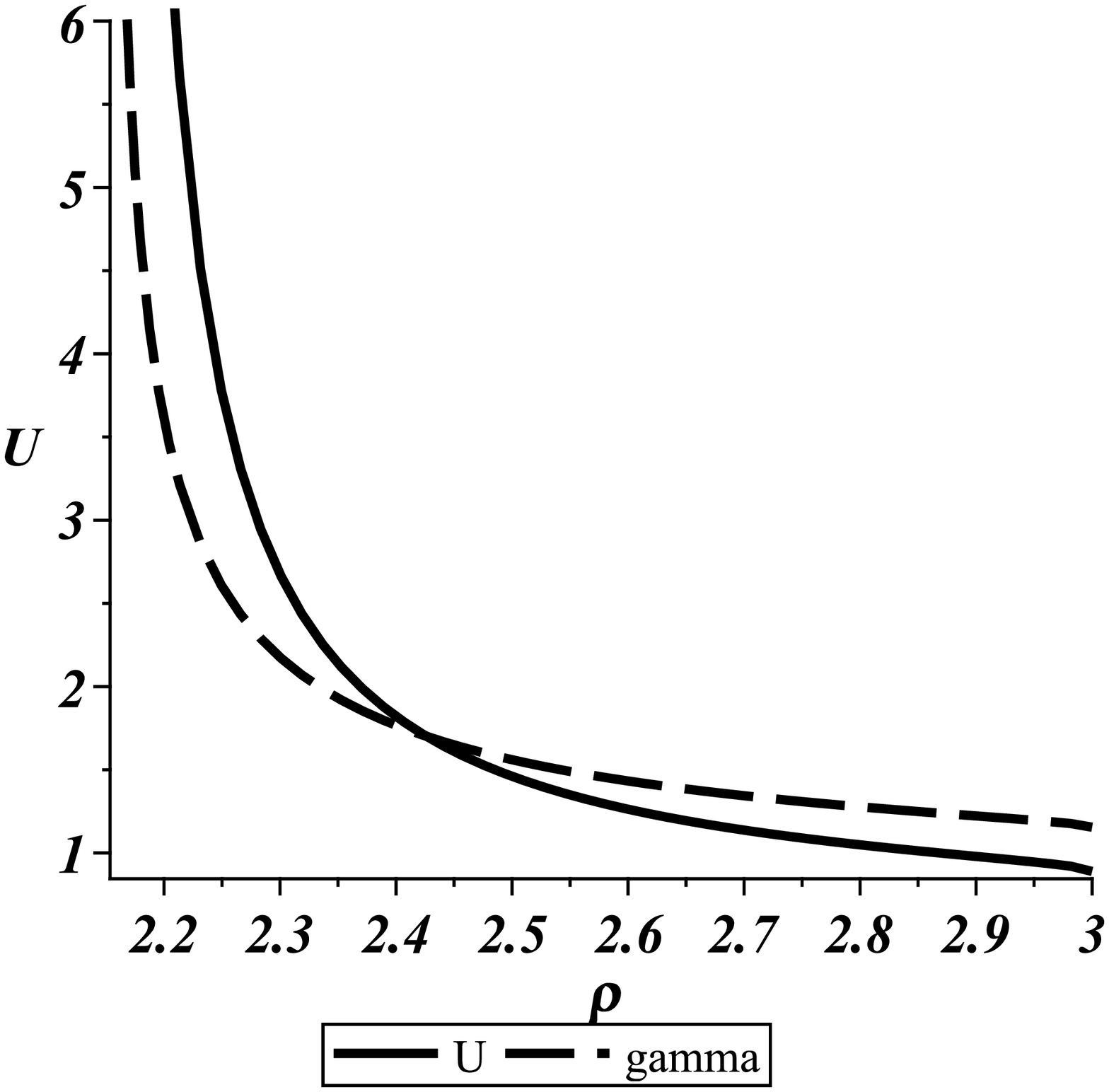}
\caption{The value of the effective potential $U_-$ at the position of the ISCO $\rho_-$ and the corresponding value of the $\gamma$-factor $\gamma_-$.}\label{fig_2m}
\end{center}
\end{figure}

Figure~\ref{fig_2p} shows that  $U_+$ monotonically decreases from its value $8/9$ at $\rho_+=3$ till $0$ at the horizon $\rho_+=1$, while the $\gamma$-factor remain always slightly higher than 1. At the end points of the ISCO radius domain one has
\be\n{ghor}
\gamma_+(1)=\gamma_+(3)={2/\sqrt{3}}\, .
 \ee
 For $b\gg 1$ the ISCO radius $\rho_+$ is close to the horizon and one has (see \cite{FS})
 \be\n{rpp}
\rho_+-1\approx {1\over \sqrt{3}b}\, .
\ee
For the same mass of the black hole $M$ and the value of the magnetic field $H$, the parameter $b$ for electrons is almost 2,000 times larger than for protons, so that the ISCO for electrons are located much closer to the horizon than the corresponding orbits for protons. Because the signes of charges for the electron and proton are different, they move along their ISCOs in the opposite direction.

 Figure~\ref{fig_2m} shows that  for trajectories with $\ell<0$ both $U_-$ and $\gamma_-$ infinitely grow at $\rho_{-,min}=(5+\sqrt{13})/4$.

\section{Particle collision}

\subsection{Two particles at ISCO}

As a first example let us consider a collision of two particles with the same mass $m$ and opposite charges $+|q|$ and $-|q|$ moving along the same circular orbit in the opposite directions. The 4-momentum of the system after the collision is $P^{\mu}=2m \gamma e_{(t)}^{\mu}$. Denote by ${\cal M}$ the energy after the collision calculated in the center-of-mass frame, then one has
\be
{\cal M}=2m\gamma\, .
\ee
Let us assume that for chosen value of the magnetic field the particles moves at the ISCO.
As one can see from Figure~\ref{fig_2p}, the value  $\gamma_+$ for an arbitrary field $b$ remains close to 1. This means that the collision energy ${\cal M}$ cannot be much higher than $2m$.

For the `attraction' case the situation is quite different. Formally both $\gamma_-$ and ${\cal M}$ infinitely grow when the radius of the ISCO tends to its minimal value $\rho_{-,min}=(5+\sqrt{13})/4$. Thus in such a collision one can obtain an arbitrary large value of ${\cal M}$ (see Figure~\ref{fig_2m}).  The energy of such particles as measured at infinity, $E=mU_{-}(\rho_-)$,  also grows in this limit. This means that in such a setup the high collision energy is possible only if the initial energy of the particles measured at infinity is also high. In other words, the gravitational field of the black hole simply `helps' the magnetic field to keep the charged particles at the circular orbit, but it does not provide  the colliding  particles with the energy.

\subsection{Collision of a freely falling  neutral particle with a charged particle at ISCO}

\subsubsection{Collision energy}

Let us consider now another case, when a freely falling from infinity  neutral particle collides with a charged particle revolving at the circular orbit near a weakly magnetized black hole. As earlier we denote by $\bm{p}$ the momentum of this particle, and by $m$ and $q$ its mass and charge. Denote by $\mu$  the mass of a freely falling particle, and by $\bm{k}$ its 4-momentum. At the moment of collision the 4-momentum  is
\be
\bm{P}=\bm{p}+\bm{k}\, ,
\ee
and the corresponding center-of-mass energy is ${\cal M}$,
\be
{\cal M}^2=m^2+\mu^2-2 (\bm{p},\bm{k})\, .
\ee

 \subsubsection{Freely falling particle}

 To calculate $(\bm{p},\bm{k})\equiv g_{\mu\nu}p^{\mu}k^{\nu}$ at the point of the collision we obtain first an expression for $\bm{k}$ in terms of the integrals of motion for the neutral particle.
We parameterize the particle geodesic by an affine parameter $\lambda$, such that
\be\n{kkk}
\bm{k}=(\dot{t},\dot{r},\dot{\theta},\dot{\phi})\hh \dot{(...)}=d(...)/d\lambda\, .
\ee
For this parametrization one has
\be
g_{tt}\dot{t}^2+g_{rr}\dot{r}^2+g_{\theta\theta}\dot{\theta}^2+g_{\phi\phi}\dot{\phi}=-\mu^2\, ,
\ee
and  the case of a massless particle (`photon') does not require  any modifications.
We use the following integrals of motion: the energy $\mathfrak{E}$, the azimuthal angular momentum $\mathfrak{L}_z$ and the total angular momentum $\mathfrak{L}$
\be
\mathfrak{E}=f\dot{t}\hhh \mathfrak{L}_z=r^2\sin^2\theta \dot{\phi}\hhh
\mathfrak{L}^2=r^4 ( \dot{\theta}^2+\sin^2\theta \dot{\phi}^2) \, .
\ee
Thus we have
\ba
\bm{k}&=&(\mathfrak{E}/f,\dot{r},\dot{\theta},\mathfrak{L}_z/r^2)\, ,\\
\dot{r}&=& \pm\sqrt{\mathfrak{E}^2-(\mu^2+\mathfrak{L}^2/r^2)f}\, ,\n{rrr}\\
\dot{\theta}&=&\pm r^2 \sqrt{\mathfrak{L}^2-\mathfrak{L}_z^2/\sin^2\theta}\n{thth}\, .
\ea

\begin{figure}[htb]
\begin{center}
\includegraphics[width=6cm]{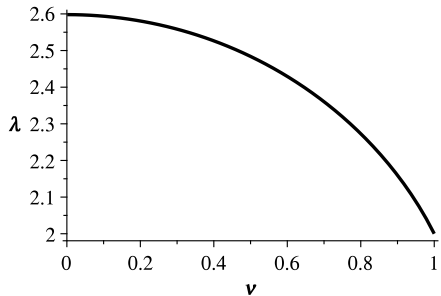}
\caption{The critical angular momentum $\lambda=\mathfrak{L}/(\mathfrak{E} r_g)$ as a function of $\nu=\mu/\mathfrak{E}$.}\label{fig_3}
\end{center}
\end{figure}

For the motion from infinity the particle is either captured by the black hole, or, after close encounter, returns to infinity. To determine the critical value of the angular momentum $\mathfrak{L}$ which separates these two outcomes we use the relation \eq{rrr} written in the form
\be
\dot{r}^2=\mathfrak{E}^2 W\hh
W=1-(\nu^2+l^2/\rho^2)(1-1/\rho)\, .
\ee
Here $\nu=\mu/\mathfrak{E}$ and $l=\mathfrak{L}/(\mathfrak{E} r_g)$.  For the motion from the infinity $\nu\le 1$, and $\nu=0$ for the ultrarelativistic particles and light. The function $W$ vanishes at the horizon and has the value $1-\nu^2$ at the infinity. Depending on the parameters $\nu$ and $l$ it is either monotonically increasing, or it has one maximum, where $dW/d\rho=0$.
The condition $W=0$ determines the radial turning points. The condition that the radial point coincides with the maximum of $W$,
\be\n{WWd}
W=dW/d\rho=0\, ,
\ee
determines the critical impact parameter $l_{crit}=\lambda(\nu)$. The capture takes place when $|l|<l_{crit}$. The function $\lambda(\nu)$ obtained by solving \eq{WWd} is shown at Figure~\ref{fig_3}.
One can see from this plot that  the critical impact parameter $\lambda$ monotonically changes from $2$ for
a particle with zero velocity at the infinity, till its maximal value $\lambda=3\sqrt{3}/2$ for ultrarelativistic particles.

\subsubsection{Properties of the collision energy}

Using this representation \eq{kkk} for $\bm{k}$ and \eq{pp} one  finds
\be
(\bm{p},\bm{k})=-m\gamma \mathfrak{E}\left(f^{-1/2}-{vl_z\over \rho}\right)\, .
\ee
where $l_z=\mathfrak{L}_z/(\mathfrak{E} r_g)$.
Thus the collision energy ${\cal M}$ obeys the equation
\be\n{MMM}
{\cal M}^2=m^2+\mu^2+2m\gamma \mathfrak{E}\left(f^{-1/2}-{vl_z\over \rho}\right)\, .
\ee

As we already know for ISCO with $\ell<0$ the $\gamma$-factor infinitely grows at the minimal radius $\rho_{-,min}$. As a result, ${\cal M}$ can be arbitrary large. But as earlier this case is not interesting for our purposes since the corresponding high energy is required from the very beginning in order to put a charged particle at such an orbit.

The `repulsion' case is more interesting. Since the radius of the ISCO can be arbitrary close to the horizon, the factor $f^{-1/2}$ in \eq{MMM} can be made arbitrary large, while $\gamma_+$ remains finite.
The second term in the brackets in \eq{MMM} remains finite. Really, from \eq{thth} it follows that
$|\mathfrak{L}_z|\le \mathfrak{L}$, so that
\be
{|vl_z|\over \rho}< |l_z|\le \lambda(\nu)\le {3\sqrt{3}\over 2}\, .
\ee
For the fixed energy $\mathfrak{E}$ the maximal value of the parameter $|vl_z|/\rho={3\sqrt{3}\over 2} $
is reached for massless particles (photons) propagating in the equatorial plane. This contribution to ${\cal M}^2$ is positive when the photon and the particle move in the  opposite direction.
In any case, this term remains finite in the limit $\rho_+\to 1$.

Thus for the collision close to the horizon the leading contribution to ${\cal M}$  is
\be
{\cal M}\sim {(2m\gamma_+ \mathfrak{E})^{1/2}\over (\rho_+-1)^{1/4}}\, .
\ee
Using relation \eq{rpp} one obtains the following asymptotic value of the center-off-mass energy for a collision with the charged particle at ISCO in a magnetic field $b\gg 1$
\be\n{Mb}
{\cal M}\sim (3)^{1/8} b^{1/4} (2m\gamma_+ \mathfrak{E})^{1/2}\, .
\ee
Using expression \eq{ghor} for the value of $\gamma_+$ at the horizon one can write \eq{Mb} in the form
\be\n{Mc}
{\cal M}\sim \alpha b^{1/4}\sqrt{m  \mathfrak{E}}\hh
\alpha={2\over 3^{1/8}}\approx 1.74\, .
\ee

In a special case, when a neutral particle falling from infinity also has the mass $m$, and it starts its motion with zero velocity one has $\mathfrak{E}=m$ and \eq{Mb}  takes a simpler form
\be\n{MMb}
{{\cal M}\over m}\sim 1.74 b^{1/4}\, .
\ee
Let us remind that for near extremal rotating black holes
the maximal collision energy per unit mass for particles close the horizon
 is (cf \cite{Jacobson:2009zg}, Eq. (8))
\be\n{MMa}
{{\cal M}\over m}\sim 4.06 (1-a)^{-1/4}
\ee
By comparing  relations (\ref{MMb}) and (\ref{MMa}) one can see that
they  are quite similar if one identifies $1-a$ with $b^{-1}$.

\section{Discussion}

We demonstrated that the collision of two charged particles moving in the magnetized black hole at the same ISCO trajectory close to the horizon in the opposite directions does not result in the high collision energy. On the other hand, this energy can be high for a collision of a particle falling from infinity with
a charged  particle revolving at ISCO. In fact, this energy formally  infinitely grows for ISCO arbitrary close to the horizon. The closeness  of ISCO to the horizon is controlled by the value of the magnetic field \eq{rpp}.
However the collision energy $M$ grows  rather slowly with the magnetic field (see \eq{MMb}). This results in a considerable suppression of the maximal collision energy for realistic  magnetic field.

To estimate the maximal collision energy per a unit mass, \eq{MMb},  one can use expression \eq{bdef} for the parameter $b$. This gives
$1.74 b^{1/4}\approx 942.3$ for a stellar mass black hole with parameters \eq{BBbesta}. For the supermassive black hole with parameters
\eq{BBbestb} this factor is of one order larger. One can expect that there exist other effects, which  restrict the ability of magnetized black hole to work as particle accelerators. One of them is the synchrotron radiation by charged particles near black holes, studied in \cite{GP,AG}. It should be also emphasized that our consideration is somehow oversimplified, since we used a free particle approximation and neglected plasma effects. 
Moreover we discussed only a simple case of ISCO particles, while the motion of charged particles
in the magnetized black holes can be more complicated (see e.g. \cite{FS}).
We conclude by the following remarks: Theoretical search for high energy events in the black hole vicinity is very interesting and important for possible astrophysical applications. In this connection, it might be interesting to study collision energy near the horizon of the black hole in the presence of both, rotation and magnetic field.

\begin{acknowledgments}
This paper was written during the author's stay at the Yukawa Institute for Theoretical Physics (YITP). I would like to thank YITP for its hospitality. The author also thanks the Natural Sciences and Engineering Research
Council of Canada and the Killam Trust for their financial support.
\end{acknowledgments}

\end{document}